# Mn-deficient Ni–Mn–Ga alloys: phase transitions, magnetic properties, and disappearance of martensitic transformation

V. Moskvina[1,2*], J. Čapek[1], M. Rameš[1], D. Musiienko[1], P. Veřtát[1], A. Perevertov[1], L. Straka[1], O. Heczko[1]

[1] *FZU – Institute of Physics of the Czech Academy of Sciences, Prague, Czech Republic*

[2] *Faculty of Nuclear Sciences and Physical Engineering, Czech Technical University in Prague, Prague, Czech Republic*

***Corresponding author:** Valentina Moskvina, FZU – Institute of Physics of the Czech Academy of Sciences, Na Slovance 1999/2, 182 00 Prague 8, Czech Republic. Email: moskvina@fzu.cz*

***Abstract***

*Composition gradients formed during directional solidification are generally considered undesirable in Ni–Mn–Ga magnetic shape-memory alloys. However, these gradients offer a highly effective platform for systematically investigating composition–property relationships under identical growth and heat-treatment conditions. To exploit this approach, one Mn-deficient and two near-stoichiometric Ni–Mn–Ga oligocrystals were grown by optical zone remelting and characterized by X-ray fluorescence mapping, optical microscopy, X-ray diffraction, low-field magnetic measurements, and vibrating-sample magnetometry. The ingots contained elongated grains and axial compositional gradients, primarily associated with the redistribution of Mn and Ga, whereas Ni remained relatively uniformly distributed. The martensitic and premartensitic transformation temperatures generally increased with increasing Mn content. Neither transformation was detected between 10 and 400 K in specimens containing 20.5–22.0 at.% Mn, all of which had e/a < 7.40. In contrast, both transformations were observed in the investigated specimens containing at least 22.1 at.% (e/a ≥ 7.46). This abrupt change may indicate a possible electron-concentration threshold for the occurrence of these transformations. Saturation magnetization increased with Mn content up to 25.8 at.%, reaching 96.9 A $m^2$/kg at 10 K and 71.2 A $m^2$/kg at 300 K. Post-growth annealing increased the transition temperatures and changed the low-field magnetic response. The results indicate that composition-gradient oligocrystals serve as an effective platform for investigating the relationships between composition, structure, and magnetic properties in Ni–Mn–Ga alloys with the same processing history.*

## 1. Introduction

Heusler alloys that undergo a martensitic transformation from a cubic to a low-symmetry phase have attracted particular attention, as they can exhibit the magnetic shape memory effect, as well as magnetocaloric and multicaloric effects [1–3]. These functional properties are associated with phase transformations induced by temperature changes, applied mechanical stress, or external magnetic fields [1–3]. Among magnetic shape memory (MSM) alloys, Ni–Mn–Ga remains one of the most extensively studied and technologically promising systems. Large magnetic-field-induced strains have been demonstrated in single crystals of this alloy [1, 2]. In well-ordered Ni–Mn–Ga alloys, the parent cubic phase typically has an $L2_1$ structure [4]. The martensitic transformation involves a reduction in crystal symmetry and can result in different martensitic structures, including modulated 5M (10M) and 7M (14M) phases and non-modulated (NM) tetragonal martensite, depending on the alloy composition and temperature [4, 5]. An additional premartensitic phase may also occur in certain Ni–Mn–Ga compositions at temperatures above the martensitic transformation temperature [6, 7]. From a practical standpoint, the martensite start temperature ($M_s$) and the austenite finish temperature ($A_f$) are crucial factors for the design and operating temperature range of MSM materials [8]. For this reason, understanding how chemical composition affects the transformation temperatures of these alloys remains an important research objective.

Numerous studies have investigated the dependence of the martensitic transformation temperature, the Curie temperature, and the magnetic properties of Ni–Mn–Ga alloys on chemical composition and elemental substitution [5, 8–15]. The effect of individual elements in the ternary Ni–Mn–Ga alloy depends on the compositional substitution path. At constant Mn content, an increase in Ga content decreases $M_s$ and $A_f$. At constant Ni concentration, increasing Mn content raises these temperatures, whereas at constant Ga content, substituting Mn for Ni decreases them [8, 9]. Unlike the structural transformation temperatures, which can vary considerably with chemical composition, the Curie temperature generally exhibits a more limited variation within the studied compositional range [5].

A single crystal is generally required to obtain a large MSM effect. However, synthesizing high-quality Ni–Mn–Ga single crystals is technically challenging due to the complex solidification behavior of off-stoichiometric alloys, and even the stoichiometric $Ni_2MnGa$ compound solidifies incongruently [16]. Moreover, significant chemical segregation during directional solidification can produce compositional gradients along the growth direction of the ingot. The Bridgman method has traditionally been used to produce Ni–Mn–Ga single crystals [16]. One of its main advantages is the ability to perform crystal growth in a sealed ampoule. This reduces elemental evaporation and promotes compositional stability. However, the method requires extensive preparation and typically involves relatively low growth rates. An alternative to the Bridgman method is zone melting using an optical furnace [17–19]. Controlling the growth rate and thermal gradient during zone melting can promote a more planar solid–liquid interface, thereby influencing grain selection. However, several large, competing columnar grains may develop simultaneously, resulting in an oligocrystalline structure that can serve as an intermediate step toward single-crystal growth [17, 18]. In both single crystals and oligocrystals produced by zone melting, chemical segregation may lead to longitudinal composition gradients, with variations in individual elemental concentrations reaching several atomic percent [16, 18]. Such variations are typically undesirable in Ni–Mn–Ga alloys because the structural transformation temperatures are highly sensitive to chemical composition [8, 9]. Here, we exploited the compositional gradient formed during zone remelting as an advantage to study in detail the significant effects of minor changes in chemical composition on transformation behavior and magnetic properties.

Thus, in this work, the compositional gradients formed during zone remelting in an optical furnace were used as an experimental tool. One Mn-deficient and two near-stoichiometric Ni–Mn–Ga oligocrystals were grown under identical conditions and subjected to the same heat treatment. This approach enabled a systematic investigation of the relationships between local chemical composition, transition temperatures, and magnetization. Particular attention was given to the Mn-deficient, Ga-enriched compositional region of the Ni–Mn–Ga ternary system. This composition region has received comparatively limited attention. We established detailed relationships between local composition and phase-transformation behavior under consistent experimental conditions by combining compositional mapping with structural and magnetic characterization.

## 2. Materials and Methods

High-purity elements (Ni, Mn, Ga ≥ 99.9%) were used to produce three polycrystalline Ni–Mn–Ga master alloys. The alloys were inductively melted in a Balzers VSG 02 vacuum furnace. Before the melting process, the furnace chamber was evacuated and backfilled twice with high-purity (5N) Ar. This procedure was used to remove residual oxygen and moisture. Melting was performed under Ar to reduce Mn evaporation. After complete melting, the melt was stirred inductively for 2 min to ensure chemical homogeneity. The melt was then cast into a copper mold. The resulting polycrystalline master-alloy rods had a diameter of approximately 8.5 mm and a length of 100 mm and were used for optical zone remelting. The master alloy compositions determined by X-ray fluorescence (XRF) analysis were $Ni_{49.9}Mn_{22.0}Ga_{28.1}$, $Ni_{49.9}Mn_{24.9}Ga_{25.2}$, and $Ni_{49.4}Mn_{24.7}Ga_{25.9}$ at.%.

One Mn-deficient (OC I) and two near-stoichiometric (OC II, OC III) Ni–Mn–Ga oligocrystals were then grown from polycrystalline master-alloy rods by zone remelting. Polycrystalline rods of the corresponding compositions, approximately 10 mm in length, were used as seeds. The growth was performed in an FZ-T-12000-X-VI-VP optical furnace using sapphire tubes as crucibles under an Ar atmosphere. The growth rate was 1 mm/h, the rod rotation rate was 5 rpm, and the lamp current was 32 A. Prior to zone remelting, the master-alloy rods were preheated from room temperature to the melting point over 40 min. The molten zone was stabilized before initiating crystal growth. The resulting crystals were approximately 11 mm in diameter and 75 mm in length.

After zone remelting, each crystal was longitudinally sectioned into two halves. One longitudinal half of each oligocrystal was used for compositional, structural, and magnetic characterization. A rectangular sample with dimensions of approximately 54 × 8 × 1.5 $mm^3$ was prepared from this half. To remove preparation-induced surface artifacts, the samples were mechanically ground and electropolished at 253 K using a mixture of one part 65% $HNO_3$ and three parts analytical-grade ethanol. Structural and low-field magnetic characterization was first performed on the samples in the as-grown state. The samples were then wrapped in Ta foil to prevent direct contact with the quartz ampoules. The ampoules were evacuated and backfilled with high-purity Ar to minimize Mn evaporation during annealing. Samples from all three oligocrystals underwent the same two-step annealing treatment. They were first annealed at 1273 K for 24 h to promote macroscopic homogenization, followed by a lower-temperature step at 1073 K for 4 h, close to the $B2 \rightarrow L2_1$ ordering transition. The samples remained in the furnace between the two steps. After the second step, the samples were furnace-cooled to room temperature. Structural and low-field magnetic measurements were then repeated on the annealed samples.

All three oligocrystals exhibited several elongated grains and an axial compositional gradient after optical zone remelting. The oligocrystal OC II was selected for the main set of structural and magnetic measurements (Table S1). Three representative specimens were selected from individual grains in each of OC I and OC III to extend the investigated compositional range. After longitudinal sectioning, heat treatment, and compositional analysis, specimens (~2 × 2 × 1.5 $mm^3$) were sequentially cut along the growth direction for XRD and PPMS/VSM measurements. The experimental workflow for Ni–Mn–Ga oligocrystals is illustrated in Figure 1.

XRF compositional mapping of the longitudinally sectioned crystals and local chemical composition measurements of individual specimens after annealing were performed under vacuum using an EDAX AMETEK ORBIS PC XRF spectrometer. The local chemical composition was determined from at least five measurements per specimen, and the average atomic concentrations of Ni, Mn, and Ga were calculated. Optical micrographs were obtained using a Keyence VHX-7000 digital microscope.

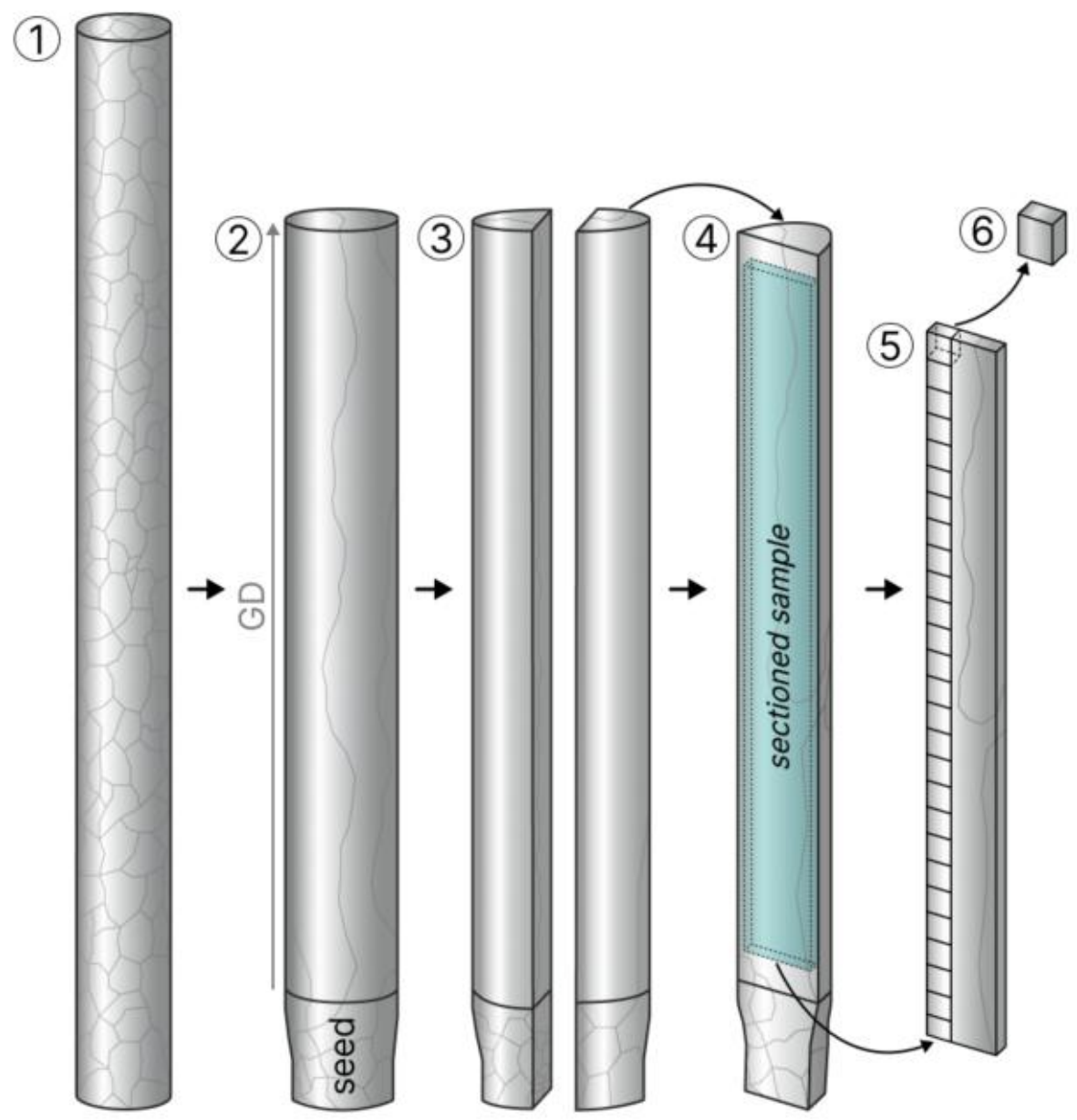

Figure 1. Schematic illustration of the experimental workflow for Ni–Mn–Ga oligocrystals: (1) polycrystalline master-alloy rod; (2) oligocrystal obtained by optical zone remelting; (3) longitudinal sectioning for XRF compositional mapping before and after annealing; (4) sectioned sample for optical microscopy, XRD and low-field AC magnetic measurements before and after annealing; (5) specimens cut sequentially along the growth direction (GD); (6) representative specimen selected from an individual grain for XRD and PPMS/VSM measurements.

X-ray diffraction (XRD) was performed at room temperature using a PANalytical X'Pert PRO diffractometer with a Co anode ($\lambda_{K\alpha 1}$ = 1.78901 Å, $\lambda_{K\alpha 2}$ = 1.79290 Å) in divergent-beam geometry. The diffractometer was equipped with an ATC-3 texture cradle. Pole figure measurements of the strong {220} and {400} reflections were performed to evaluate the sample orientation and detect any additional grains (or distinct crystallographic orientations). During the measurements, the detector acceptance angle for intensity integration was set to 2.1° in $2\theta$ to accommodate potential variations in the lattice parameter across the compositional gradient. For selected poles, asymmetric Bragg–Brentano scans were performed to estimate the lattice parameter.

Low-field AC magnetic measurements were performed using a custom single-yoke setup with a magnetically closed yoke–sample configuration. The sample was magnetized using a driving coil wound around the yoke, while a pickup coil tightly wrapped around the middle of the sample provided a signal proportional to the rate of change of magnetic flux density, *dB/dt* [20, 21]. The sample and yoke legs were polished to a mirror finish to reduce the air gaps to 10 μm or less. This ensured that the internal magnetic field in the sample was proportional to the current in the driving coil, allowing its direct determination without the demagnetization corrections typically required in conventional VSM or SQUID magnetometry. The resulting uncertainty in the differential susceptibility was below 5%, providing high sensitivity to changes in magnetization. A triangular voltage waveform was applied to the driving coil at a frequency of 0.1 Hz, producing a maximum magnetic field of 6000 A/m.

To measure the temperature dependence of magnetic properties, the entire yoke–sample assembly was immersed in a liquid nitrogen bath inside a polystyrene box. The sample temperature was monitored using a thin T-type thermocouple attached to its surface near the pickup coil. Magnetic hysteresis loops were recorded over the temperature range of 200–400 K at 10 s intervals, corresponding to approximately 0.2 K per measurement step. Further details of the measurement setup can be found elsewhere [20].

To complement the high-sensitivity low-field measurements, the magnetic and structural transition temperatures were also investigated using vibrating sample magnetometry (VSM) in a Physical Property Measurement System (PPMS) equipped with a 9 T superconducting magnet (Quantum Design, Inc., San Diego, USA). Magnetization curves were measured in magnetic fields $\mu_0 H$ up to 9 T at various temperatures in the range of 10–400 K. The saturation magnetization ($M_{sat}$) was determined from the high-field regions of the measured $M(H)$ curves.

Temperature-dependent magnetization $M(T)$ was recorded in a magnetic field of $\mu_0 H = 0.01$ T during heating and cooling cycles between 10 and 400 K at a temperature sweep rate of 4 K/min. The characteristic start and finish temperatures of the forward ($M_s$, $M_f$) and reverse ($A_s$, $A_f$) martensitic transformations were determined from the $M(T)$ curves using the tangent intersection method. The characteristic temperatures of the forward and reverse transformations were calculated as $T_M = (M_s + M_f)/2$ and $T_A = (A_s + A_f)/2$, respectively.

## 3. Results and Discussion

Zone remelting in an optical furnace produced oligocrystal ingots with compositional gradients along the growth direction. XRF mapping revealed that the axial compositional variation was mainly associated with the redistribution of Mn and Ga, whereas the Ni concentration remained comparatively homogeneous over the main growth region. For OC I, the compositional variation in this region was approximately 1.0 at.% for Ni, 2.0 at.% for Mn, and 1.5 at.% for Ga. OC III showed a less monotonic variation of approximately 1.0 at.% Ni, 3.0 at.% Mn, and 2.0 at.% Ga. The most pronounced systematic gradient was observed for ingot OC II, whose initial composition was closest to that of stoichiometric $Ni_2MnGa$ ($Ni_{49.9}Mn_{24.9}Ga_{25.2}$). The composition distribution is shown in Figure 2a. Within the section selected for further characterization, the Mn concentration increased by approximately 3.5 at.%, while the Ga concentration decreased by approximately 2.5 at.%; the corresponding change in Ni was smaller, approximately 1.0 at.%. XRF mapping was repeated after annealing and confirmed that the longitudinal compositional gradients and overall elemental distributions remained essentially unchanged. Thus, although annealing was intended to promote local chemical homogenization, it did not eliminate the macroscopic compositional gradients along the ingots.

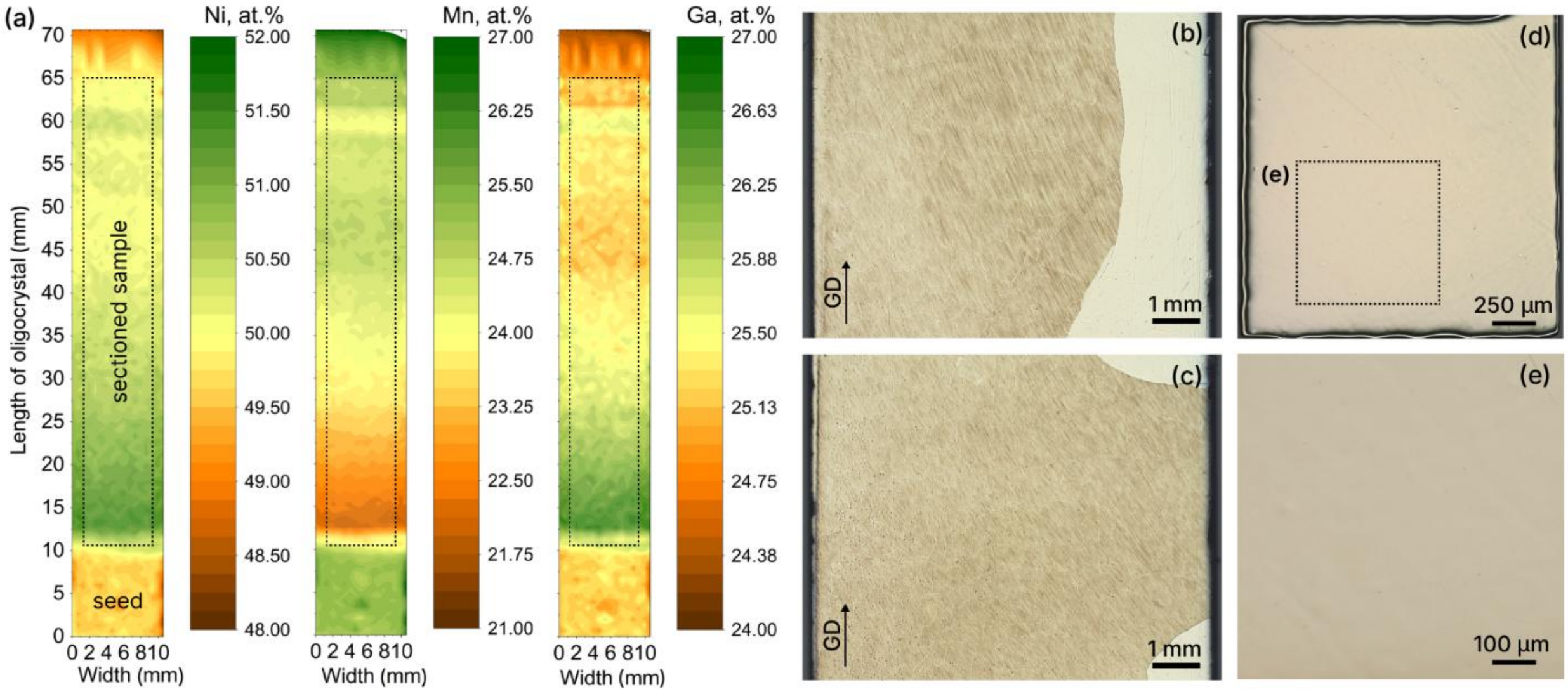

Figure 2. XRF compositional maps of Ni, Mn, and Ga in the longitudinally sectioned OC II (a); the dashed rectangle indicates the region used for sequential sample preparation. Optical micrographs of the thermally etched upper (b) and middle (c) regions, showing the grain structure along the GD. Representative specimen sectioned from an individual grain after electropolishing (d, e).

The observed axial redistribution of the constituent elements is consistent with the solidification behavior of Ni–Mn–Ga alloys. Stoichiometric $Ni_2MnGa$ has been reported to melt incongruently, and chemical segregation during directional solidification can therefore produce substantial compositional variations along the growth direction [16–18].

After the two-step heat treatment, optical microscopy revealed a coarse oligocrystalline structure with two or three elongated grains extending approximately parallel to the growth direction (GD) (Figure 2b, c). The individual grains extended over several tens of millimeters along the GD and were several millimeters wide. A similar coarse, elongated grain morphology was observed for all three oligocrystals. This is consistent with previous zone-melting studies [17, 18]. Based on these observations, individual specimens were sequentially sectioned along the growth direction within selected columnar grains. OC II was used for the main experimental dataset because it combined a well-defined longitudinal Mn/Ga gradient with sufficiently large, elongated grains. OC I and OC III were used to extend the investigated composition range.

A representative specimen cut from an individual grain is shown in Figure 2d. The enlarged optical micrograph in Figure 2e does not show any grain boundary within the investigated area, consistent with the specimen being sectioned from inside a single grain.

XRD performed in different regions of OC II confirmed an overall cubic austenite structure. Measurements at three distinct locations along the ingot confirmed its oligocrystalline nature: while the region at one location appeared to be single-crystalline, the other two exhibited up to three crystallographic orientations (grains or domains) with misorientations of up to ~20°. The corresponding asymmetric Bragg–Brentano scans indicated broadened reflections and a variation in the lattice parameter, ranging from 5.84 to 5.86 Å. Following the sample annealing, the XRD pole figures indicated subtle changes in the microstructure: the misorientation angles generally decreased (typically to 2–6°), except for one domain exhibiting a ~17° misorientation.

The representative (004) and (220) pole figures in Figure 3 show that the sample in Figure 2e consisted of a single grain, i.e., a single crystal. The sample surface normal was misoriented by approximately 13° from the [111] direction, corresponding closely to the high-index [579] direction.

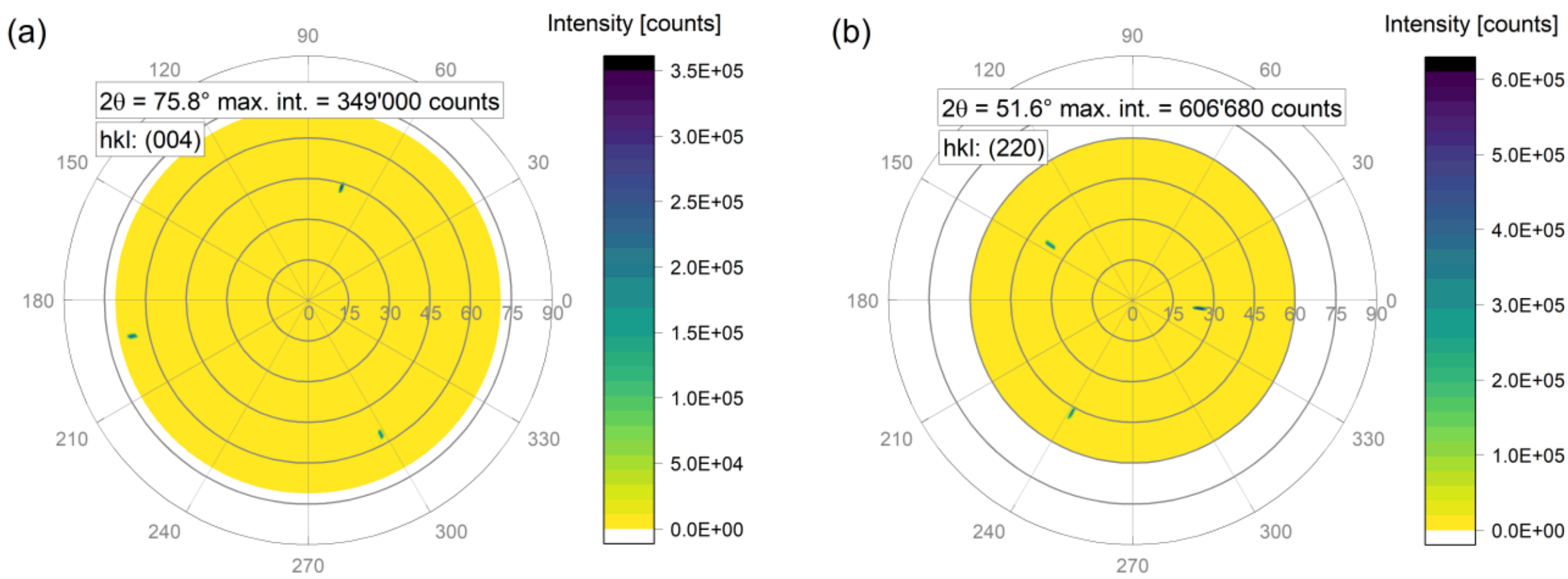

Figure 3. XRD pole figures of the (004) (a) and (220) (b) reflections of the sectioned Ni–Mn–Ga specimen in the austenitic state. The outer white areas indicate unmeasured regions. The tilt angle was limited to 70° in (a) and 60° in (b) to obtain sufficient orientation data while avoiding severe defocusing at higher angles.

To characterize the overall magnetic behavior of the ingot, the long sample was measured using the AC magnetic induction method. The magnetization at the maximum applied field ($H$ = 6000 A/m) and the maximum differential permeability as functions of temperature are shown in Figure 4. Representative magnetization curves measured at 300 K are presented in Figure S1 of the Supplementary Material. Below the martensitic transition, both magnetization and susceptibility were low due to the large magnetocrystalline anisotropy of 10M martensite [22] and the presence of differently oriented martensitic twin variants. The applied field of 6000 A/m was small compared with the anisotropy field and was therefore insufficient to approach magnetic saturation. A field approximately 200 times higher would be required to reach saturation in martensite. The sharp increase in magnetization and susceptibility indicated the transformation to cubic austenite, which is magnetically very soft [20] and can be brought close to saturation by the applied AC field. Upon further heating, the susceptibility exhibited a broad local minimum associated with the premartensitic transformation.

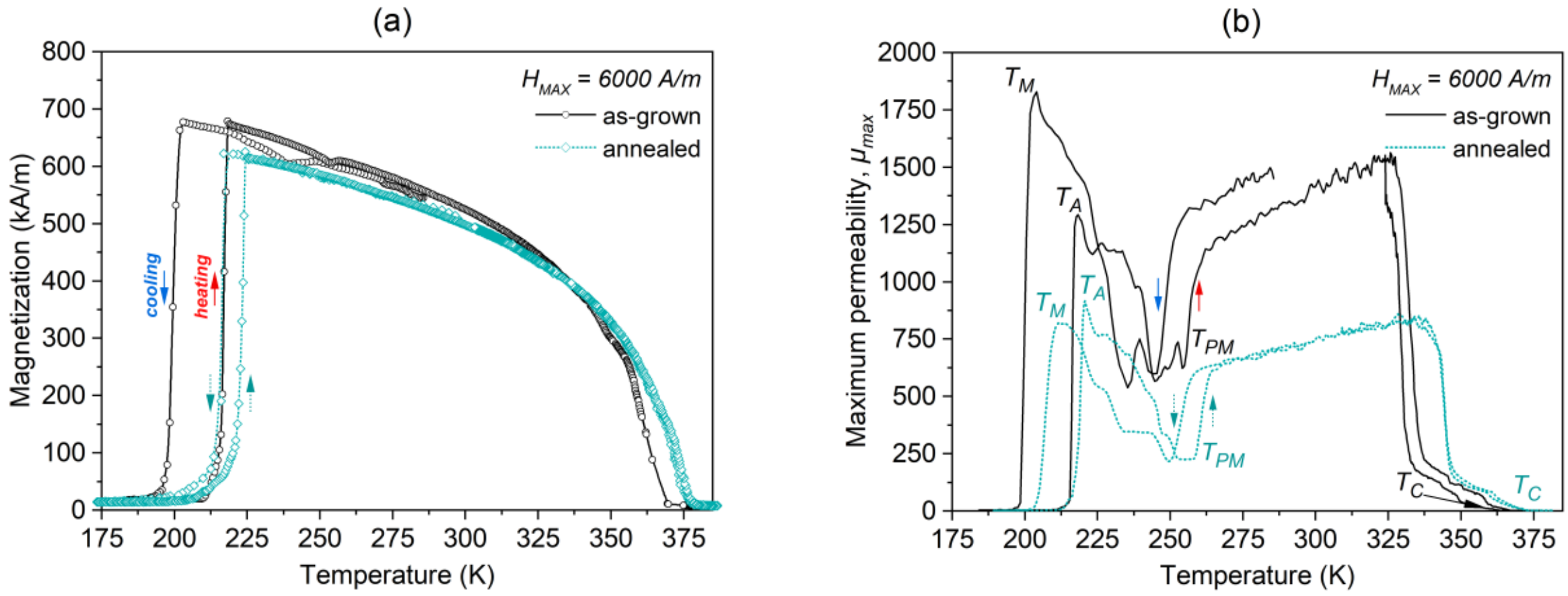

Figure 4. Temperature dependence of the maximum magnetization (a) and differential permeability (b) of the as-grown and annealed sectioned sample. The characteristic temperatures of the martensitic ($T_M$), austenitic ($T_A$), premartensitic ($T_{PM}$), and Curie ($T_C$) temperatures are indicated in (b). Arrows indicate cooling and heating; the sample was first cooled from room temperature and then heated.

As previously demonstrated in [20], the magnetic easy axis in nearly stoichiometric Ni–Mn–Ga rotates by 45° after the reverse martensitic transformation, accompanied by an increase in magnetic anisotropy. Consequently, rather than continuing to increase, the susceptibility began to decrease immediately after the transformation to austenite, reaching a minimum near the premartensitic transition temperature of 254 K. At higher temperatures, the magnetic easy axis rotated back towards [100], the measurement direction, and the susceptibility increased again. However, due to the significant compositional gradient along the ingot, the premartensitic transformation occurred at different temperatures across the sample, resulting in a broadened susceptibility minimum. A small tail was also observed in the susceptibility curve just below the Curie temperature, at which both susceptibility and magnetization decreased sharply. In contrast to its pronounced effect on susceptibility, the premartensitic transformation has little influence on the saturation magnetization, whose temperature dependence can be described by a classical Lamé curve of *M(T)* [21]. After annealing, the characteristic transformation temperatures shifted noticeably towards higher values.

To determine the saturation magnetization and characterize the magnetic behavior of selected compositions, measurements were performed on individual specimens prepared as described in the Materials and Methods section. Figure 5a shows the temperature dependence of magnetization measured in a low field of 0.01 T for three representative compositions: $Ni_{50.1}Mn_{22.0}Ga_{27.9}$ (*e*/*a* = 7.39), $Ni_{51.1}Mn_{22.1}Ga_{26.8}$ (*e*/*a* = 7.46), and $Ni_{50.3}Mn_{24.7}Ga_{25.0}$ (*e*/*a* = 7.51).

All three alloys were ferromagnetic at room temperature. The transition to the paramagnetic state appeared as a sharp decrease in *M(T)* at the Curie temperature, $T_C$, which ranged from 340 to 380 K for these compositions. However, only two of the three selected compositions exhibited martensitic and premartensitic transformations, as indicated by the *M(T)* curves. The weak anomaly at about 218–266 K (Figure 5a, Table S1) was attributed to the premartensitic transition ($T_{PM}$); the significant, abrupt change in low-field magnetization near 183–225 K corresponded to the forward and reverse martensitic transformations. The hysteresis of the first-order martensitic transformation was marked by a distinct divergence between the cooling and heating curves.

Despite the similarity of the first alloy's composition to that of the second alloy, its *M(T)* curve did not exhibit any signature of a martensitic or premartensitic transition within the 10–400 K range; the magnetization remained nearly constant until it dropped sharply at $T_C$. The absence of a structural transformation was unexpected, because its composition was very similar to that of the other alloys (Table S1). This may be associated with an *e*/*a* ratio below 7.40. This value is consistent with predictions for other Heusler alloys. The transition appears to be highly sensitive to *e*/*a*: the alloys with *e*/*a* ratios of 7.51 and 7.46 exhibited martensitic transformations at 214 and 183 K, respectively, whereas no transformation was detected for the alloy with *e*/*a* = 7.39.

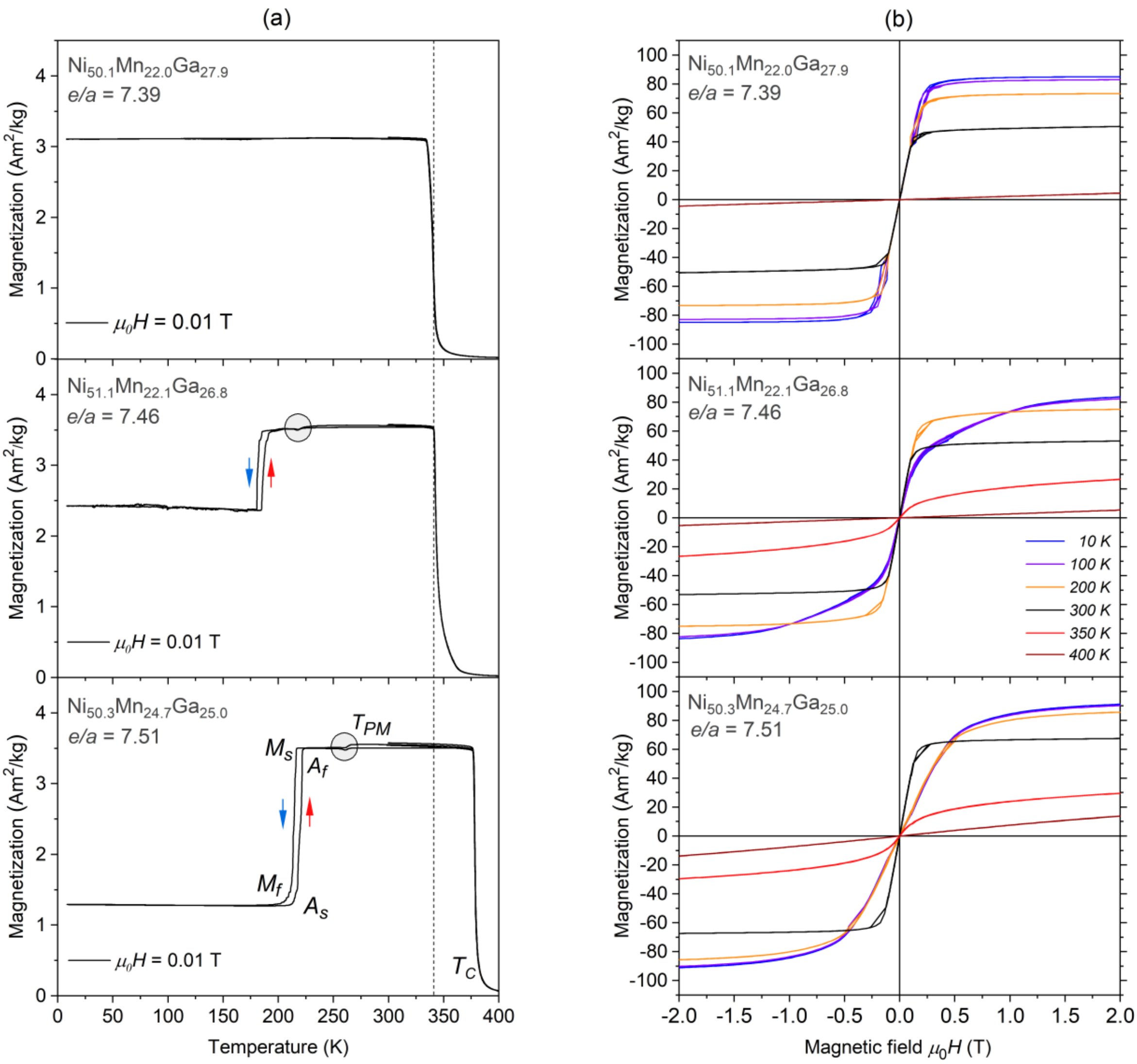

Figure 5. Temperature dependence of magnetization, measured at $\mu_0 H$ = 0.01 T, with indicated characteristic transformation temperatures (a), and magnetic hysteresis loops *M(H)* measured at selected temperatures (b) for Ni–Mn–Ga specimens with different compositions.

The hysteresis loops in Figure 5b indicated the soft-magnetic ferromagnetic nature of austenite, specifically its low coercivity and narrow hysteresis loop in the central region. At 350 K, the curves remained nonlinear and did not reach full saturation at 2 T due to the proximity to $T_C$, which increased thermal fluctuations in the magnetic order. At 400 K, the *M(H)* curves were nearly linear with a low slope, characteristic of the paramagnetic state above $T_C$. In martensite, the magnetization curves at 10 and 100 K showed a slower approach to saturation and an extended high-field region. This behavior was attributed to the high magnetocrystalline anisotropy of martensite and the existence of differently oriented twin variants.

The saturation magnetization and transformation temperatures were determined from magnetization curves at 10 K and 300 K and from thermomagnetic curves measured for the entire set of sectioned samples. The variation in saturation magnetization and characteristic transition temperatures as a function of Mn content is shown in Figure 6.

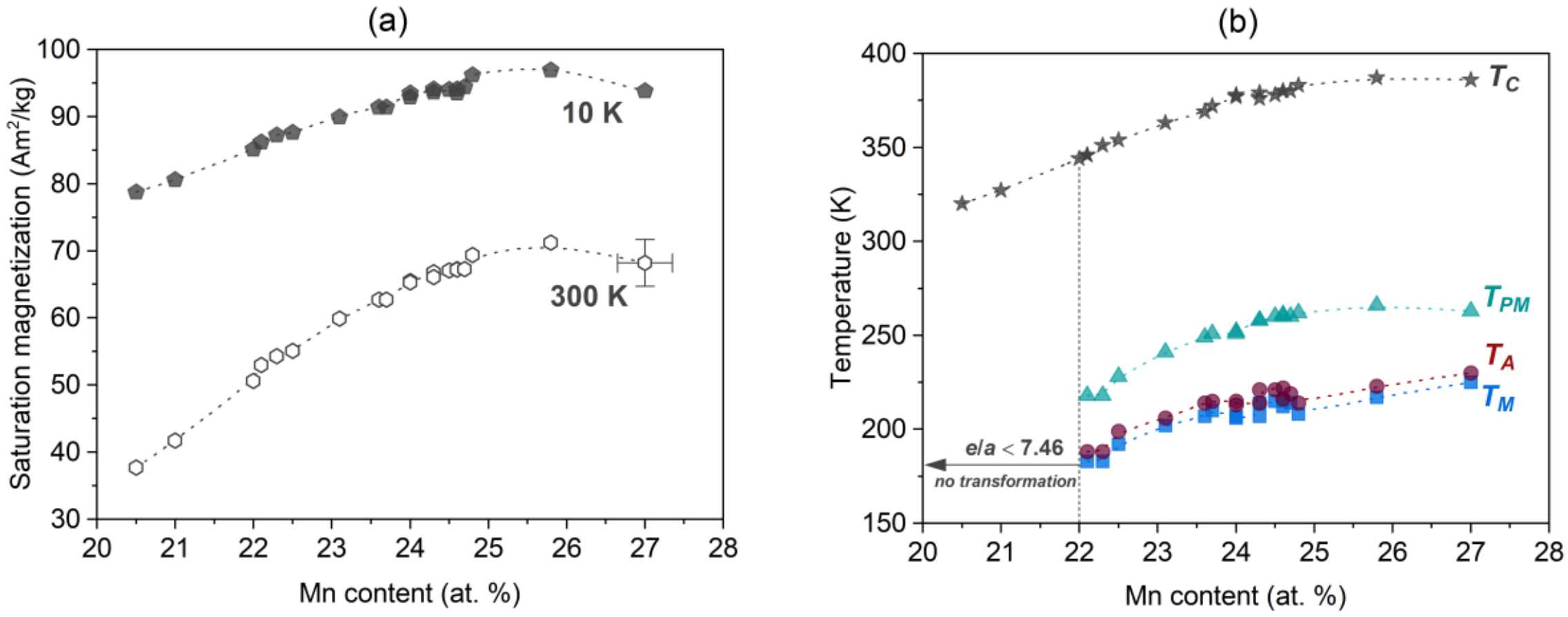

Figure 6. Saturation magnetization ($M_{sat}$) at 10 and 300 K (a) and characteristic transition temperatures (b) as functions of Mn content. The martensitic, austenitic, premartensitic, and Curie temperatures are denoted as $T_M$, $T_A$, $T_{PM}$, and $T_C$, respectively. The lines are guides for the eyes.

Saturation magnetization at 10 K increased from approximately 78 A m$^2$/kg at 20.5 at.% Mn to a maximum of approximately 97 A m$^2$/kg near 25–26 at.% Mn and subsequently decreased to approximately 93 A m$^2$/kg at 27 at.% Mn (Figure 6a, Table S1). The maximum near the stoichiometric composition may reflect favorable ferromagnetic ordering at stoichiometry (full $L2_1$ order), and any deviations from $Ni_2MnGa$ reduce the net magnetic moment. In Mn-rich alloys (excess Mn atoms), the decrease is attributed to antiferromagnetically coupled excess Mn moments [23]. In Mn-deficient alloys, the magnetization decreased approximately linearly with the number of magnetic Mn atoms. Indeed, a linear extrapolation of our data to zero Mn gave zero magnetization. This suggests that the magnetic moment of Mn is not strongly affected by the surroundings, at least within the investigated composition range.

At room temperature (300 K), the compositional dependence of $M_{sat}$ showed a similar maximum, but the absolute values were lower, ranging from approximately 37 to 71 A m$^2$/kg (Figure 6a, Table S1). The decrease in magnetization compared with that at 10 K results from the reduction in spontaneous magnetization with increasing temperature; the relative change depends on the distance to the Curie point [21].

As shown in Figure 6b, the transformation temperatures $T_M$ and $T_A$ increased with Mn content, while the thermal hysteresis of the martensitic transformation remained approximately constant. The characteristic temperature $T_M$ rose from approximately 183 K to 208–215 K over the range of 22.1–25 at.% Mn. With a further increase in Mn content to 26–27 at.%, $T_M$ reached 225 K (Table S1). In the Mn concentration range of 24.5–25 at.%, slight fluctuations in $T_M$ were observed. Since Mn enrichment along the investigated compositional path was accompanied primarily by Ga depletion, this trend should be interpreted as an effect of the coupled increase in Mn content and decrease in Ga content, rather than as an independent single-parameter influence of Mn.

The premartensitic transition temperature, $T_{PM}$, showed a similar overall trend and increased with Mn concentration. It ranged from approximately 218–220 K at about 22 at.% Mn to 250–260 K at 24–25 at.% Mn. At higher Mn contents, $T_{PM}$ approached a plateau at approximately 260–266 K. Indeed, for a typical alloy with a martensitic transformation temperature close to room temperature, the premartensitic transition is not detectable [25].

The Curie temperature, $T_C$, exhibited a trend similar to that of $T_{PM}$ as the Mn concentration increased to 27 at.%. It rose from approximately 320 K at 20.5 at.% Mn to 387 K near 26 at.% Mn and then remained nearly constant (Figure 6b, Table S1). Like saturation magnetization, the Curie temperature in Mn-deficient alloys decreased nearly linearly with decreasing Mn content, suggesting a weakening of the interaction between the remaining magnetic Mn atoms.

As discussed above, the most interesting point that can be inferred from Figure 6b is the disappearance of the martensitic transformation in alloys with $e/a < 7.40$ within the investigated temperature range. The alloys with $e/a \geq 7.46$ exhibited relatively high $T_M$, which decreased approximately linearly with decreasing Mn content. However, no transformation was detected between 10 and 400 K in the three alloys with $e/a < 7.40$ and similar Mn contents. These observations suggest a possible threshold in electron concentration for the martensitic transformation, consistent with previous suggestions and analyses of various Heusler alloys [25, 26]. Although the martensitic transformation disappeared, the Curie temperature remained measurable.

## 4. Conclusions

Ni–Mn–Ga oligocrystals containing large, elongated grains were produced by zone remelting in an optical furnace. The ingots exhibited axial Mn/Ga compositional gradients that enabled systematic investigation of composition-dependent structural and magnetic transitions.

In the Mn-deficient crystals, the premartensitic and martensitic transformation temperatures decreased approximately linearly with decreasing Mn content, accompanied primarily by increasing Ga content. Transitions were observed from 22.1 at.% Mn upward and for $e/a$ ratios of at least 7.46. However, no premartensitic or martensitic transition was detected down to 10 K in the 20.5–22.0 at.% Mn range for $e/a < 7.40$.

Saturation magnetization showed a maximum near 25–26 at.% Mn at both 10 and 300 K. In the Mn-deficient alloys, saturation magnetization and the Curie temperature decreased approximately linearly with decreasing Mn content. The decrease in the Curie temperature was consistent with weaker magnetic interactions.

Post-growth annealing shifted the characteristic transformation temperatures to higher values and changed the low-field magnetic response. Thermal history should therefore be considered when comparing transformation temperatures of Ni–Mn–Ga alloys with similar compositions.

We demonstrated that oligocrystalline ingots containing large, elongated grains with well-characterized compositional gradients provided a highly suitable platform for systematic composition–property studies, enabling several closely related compositions to be compared under the same processing history.

**Declaration of Competing Interest**

The authors declare that they have no known competing financial interests or personal relationships that could have appeared to influence the work reported in this paper.

**Acknowledgement**

We acknowledge the support provided by the Ferroic Multifunctionalities project (FerrMion), supported by the Ministry of Education, Youth, and Sports of the Czech Republic [OP JAK project No. CZ.02.01.01/00/22_008/0004591], co-funded by the European Union, and the Czech Science Foundation [Project No. 23-04806S]. Magnetic measurements were performed at the MGML (http://mgml.eu), supported by the program of Czech Research Infrastructures [Project No. LM2023065].

**Data availability**

The data supporting the findings of this study are available [27].

**Declaration of generative AI and AI-assisted technologies in the manuscript preparation process**

During the preparation of this work, the authors used ChatGPT (OpenAI) to improve the manuscript's readability and language. After using this tool, the authors reviewed and edited the content as needed and take full responsibility for the content of the publication.

# Supplementary Material

## Mn-deficient Ni–Mn–Ga alloys: phase transitions, magnetic properties, and disappearance of martensitic transformation

V. Moskvina[1,2*], J. Čapek[1], M. Rameš[1], D. Musiienko[1], P. Veřtát[1], A. Perevertov[1], L. Straka[1], O. Heczko[1]
[1] *FZU – Institute of Physics of the Czech Academy of Sciences, Prague, Czech Republic*
[2] *Faculty of Nuclear Sciences and Physical Engineering, Czech Technical University in Prague, Prague, Czech Republic*

Table S1. Local chemical composition, characteristic transition temperatures, and saturation magnetization of specimens obtained from three Ni–Mn–Ga oligocrystals.

| Oligocrystal | Ni | Mn | Ga | $e/a$ | $T_M$ | $T_A$ | $T_{PM}$ | $T_C$ | $M_{sat}^{10K}$ | $M_{sat}^{300K}$ |
|---|---|---|---|---|---|---|---|---|---|---|
| | at.% | | | – | K | | | | A m²/kg | |
| OC I | 50.8 | 20.5 | 28.7 | 7.38 | – | – | – | 320 | 78.7 | 37.7 |
| | 50.4 | 21.0 | 28.6 | 7.37 | – | – | – | 327 | 80.6 | 41.7 |
| | 50.1 | 22.0 | 27.9 | 7.39 | – | – | – | 344 | 85.1 | 50.6 |
| OC II | 51.1 | 22.1 | 26.8 | 7.46 | 183 | 188 | 218 | 346 | 86.2 | 53.0 |
| | 51.1 | 22.3 | 26.6 | 7.47 | 183 | 188 | 218 | 351 | 87.2 | 54.3 |
| | 51.1 | 22.5 | 26.4 | 7.48 | 192 | 199 | 228 | 354 | 87.6 | 55.1 |
| | 51.0 | 23.1 | 25.9 | 7.49 | 202 | 206 | 241 | 363 | 89.9 | 59.9 |
| | 50.5 | 23.6 | 25.9 | 7.48 | 207 | 214 | 249 | 369 | 91.4 | 62.7 |
| | 50.8 | 23.7 | 25.5 | 7.50 | 210 | 215 | 251 | 372 | 91.4 | 62.7 |
| | 50.5 | 24.0 | 25.5 | 7.50 | 207 | 215 | 251 | 377 | 93.5 | 65.5 |
| | 50.4 | 24.0 | 25.6 | 7.49 | 206 | 213 | 252 | 378 | 92.9 | 65.3 |
| | 50.4 | 24.3 | 25.3 | 7.50 | 207 | 214 | 258 | 379 | 94.1 | 66.8 |
| | 50.3 | 24.3 | 25.4 | 7.49 | 214 | 221 | 258 | 376 | 93.6 | 66.1 |
| | 50.2 | 24.6 | 25.2 | 7.50 | 215 | 222 | 260 | 380 | 94.1 | 67.3 |
| | 50.2 | 24.5 | 25.3 | 7.49 | 215 | 221 | 260 | 378 | 94.0 | 67.1 |
| | 50.1 | 24.6 | 25.3 | 7.49 | 212 | 216 | 261 | 380 | 93.5 | 67.2 |
| | 50.3 | 24.7 | 25.0 | 7.51 | 214 | 219 | 260 | 380 | 94.1 | 67.3 |
| OC III | 50.0 | 24.8 | 25.2 | 7.49 | 208 | 214 | 262 | 383 | 96.6 | 69.4 |
| | 49.5 | 25.8 | 24.7 | 7.50 | 217 | 223 | 266 | 387 | 96.9 | 71.2 |
| | 49.0 | 27.0 | 24.0 | 7.51 | 225 | 230 | 263 | 386 | 93.8 | 68.2 |

The valence electron concentration was calculated as $e/a = (10c_{Ni} + 7c_{Mn} + 3c_{Ga})/100$, using 10, 7, and 3 valence electrons for Ni, Mn, and Ga, respectively [1].

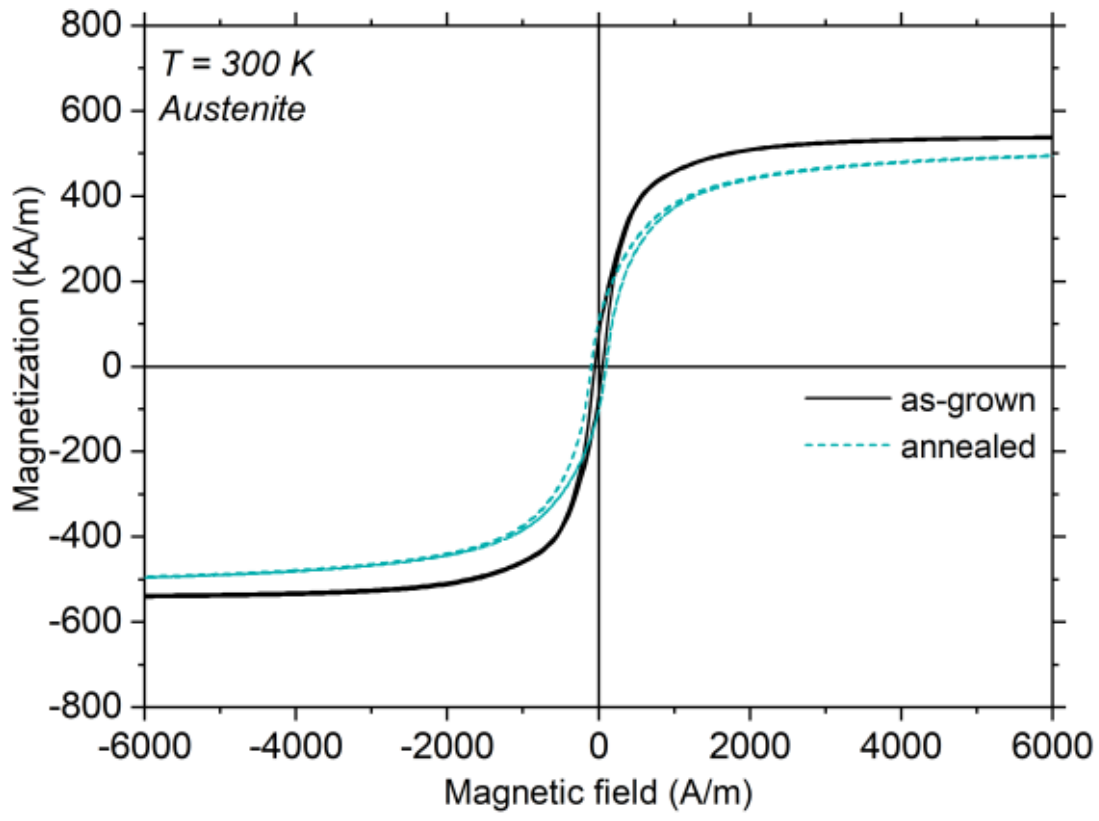

Figure S1. Magnetization curves of the as-grown and annealed Ni–Mn–Ga sectioned sample measured at 300 K in the austenitic state using the AC method. The magnetization at the maximum applied field and the maximum differential permeability were determined from these curves. The data complement those presented in Figure 4 of the main text.